\UseRawInputEncoding
\documentclass[12pt]{iopart}

\usepackage{graphicx}
\usepackage{subcaption}
\usepackage{multirow}
\usepackage{amssymb}
\usepackage{bm}
\usepackage{IEEEtrantools}

\begin{document}

\title{Towards Transparent and Accurate Plasma State Monitoring at JET}

\author{A. B\"urli$^1$, A. Pau$^2$,
T. Koller$^1$ and O. Sauter$^2$ and JET Contributors$^3$ }
\address{$^1$ Department of Computer Science, Lucerne University of Applied Sciences and Arts (HSLU), 6343 Rotkreuz, CH}
\address{$^2$ Swiss Plasma Center (SPC), \'Ecole Polytechnique F\'ed\'erale de Lausanne (EPFL), 1015 Lausanne,
CH}
\address{$^3$ See the author list of \cite{10.1088/1741-4326/ad3e16}}

\ead{alessandro.pau@epfl.ch}

\vspace{10pt}

\begin{indented}
\item[]March 2024
\end{indented}

\begin{abstract}
Controlling and monitoring plasma within a tokamak device is complex and challenging. Plasma off-normal events, such as disruptions, are hindering steady-state operation. For large devices, they can even endanger the machine's integrity and it represents in general one of the most serious concerns for the exploitation of the tokamak concept for future power plants. Effective plasma state monitoring carries the potential to enable an understanding of such phenomena and their evolution which is crucial for the successful operation of tokamaks.
	
This paper presents the application of a transparent and data-driven methodology to monitor the plasma state in a tokamak. Compared to previous studies in the field, supervised and unsupervised learning techniques are combined. The dataset consisted of 520 expert-validated discharges from the ILW campaigns at JET. The goal was to provide an interpretable plasma state representation for the JET operational space. By leveraging multi-task learning for the first time in the context of plasma state monitoring and disruption prediction, a state- and a sequence-based approach was developed and compared. When evaluated as disruption predictors, a sequence-based approach showed significant improvements compared to the state-based models. The best resulting network achieved a promising cross-validated success rate when combined with a physical indicator and accounting for nearby instabilities. Furthermore, it closely matched the cumulative warning time distribution of the ground truth, showing potential for disruption prevention. Qualitative evaluations of the learned latent space uncovered operational and disruptive regions as well as patterns related to learned dynamics and global feature importance.

 The applied methodology provides novel possibilities for the definition of triggers to switch between different control scenarios, data analysis, and learning as well as exploring latent dynamics for plasma state monitoring. It also showed promising quantitative and qualitative results with warning times suitable for avoidance purposes and distributions that are consistent with known physical mechanisms.
\end{abstract}

%
%
%
%
\ioptwocol

\section{Introduction}

Phenomena such as plasma disruptions pose a serious risk to the structural integrity of tokamak devices and may also force long maintenance interventions which could significantly reduce the device availability \cite{eidietis_implementing_2018}. To mitigate or avoid disruptions, a specific and often unknown sequence of actions is required. The particular sequence of actions is defined in the complex framework of disruption prevention and exception handling, which is at the basis of modern plasma control systems. 
Therefore, the development of effective plasma state monitoring techniques is essential for further exploration, physics understanding, and improvement of the operational regions of fusion devices \cite{vu_integrated_2021}.

Supervised machine learning and more recently also deep learning methods have demonstrated remarkable capabilities in extracting and processing complex patterns from various types of data \cite{goodfellow_deep_2016}. This establishes them as promising approaches for plasma state monitoring related tasks such as disruption prevention in nuclear fusion research \cite{vega_disruption_2022, kates-harbeck_predicting_2019, ratta_advanced_2010, tinguely_application_2019, montes_machine_2019}. However, these supervised learning techniques often suffer from the so-called ``black box`` problem, where the inner workings and decision-making processes of the models are not transparent to the researchers \cite{rea_real-time_2019, pau_machine_2019}. 
To overcome these limitations, several unsupervised learning techniques have been proposed, including manifold learning techniques, such as generative topographic mapping, and variational autoencoders \cite{pau_first_2018, wei_dimensionality_2021}. These techniques aim to provide a low-dimensional representation of the operational space, allowing researchers to qualitatively inspect and understand the underlying patterns that drive disruptive events and thus enhance model transparency. 

In this work, we propose a system for plasma state monitoring at JET that integrates supervised as well as unsupervised learning techniques. We define plasma state monitoring as the process of continuously observing and analyzing plasma behavior to detect and predict changes in relevant operational parameters, as well as to further understand the mechanisms that lead to those changes. By leveraging the benefits of both learning paradigms, our approach aims to provide a meaningful representation of plasma states, identifying critical conditions and enhancing the understanding of plasma state evolution in the different operational regions. We consider both state-based and sequence-based models and emphasize their differences by evaluating them as disruption predictors.

The remaining sections are organized as follows: Section 2 provides an overview of the employed model architecture as well as the optimization scheme. Section 3 elaborates on the dataset constitution including features selection and preprocessing. Section 4 presents the quantitative results for both the state- and sequence-based approaches. Section 5 qualitatively analyses the learned low-dimensional plasma state representations and studies the projection of unseen discharges. We conclude with a summary and proposed future work.


\begin{figure*}[h]
    \centering
    \includegraphics[width=0.95\textwidth]{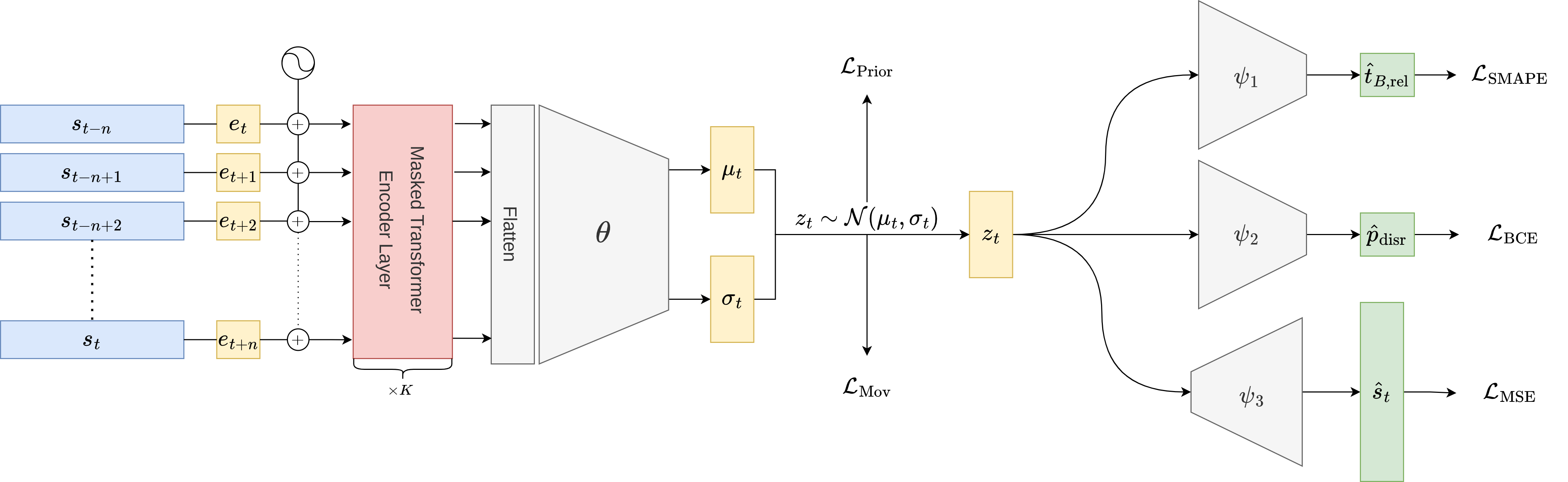}
    \caption{Sequence-based architecture of the employed model leveraging multi-task learning. By using auto-regressive attention masks the transformer encoder ensures representation and prediction causality.}
    \label{fig:architecture}
\end{figure*}

\section{Methodology}
Multitask learning (MTL) is an inductive transfer approach that holds the potential to enhance generalization by using domain-specific information in training signals of related tasks \cite{caruana_multitask_1997}. This is achieved by training tasks simultaneously while using shared representations. As a result, the training signals from related tasks can act as an inductive bias. We leverage MTL to combine supervised as well as unsupervised objectives partially inspired by prior work \cite{pau_machine_2019, kates-harbeck_predicting_2019, wei_dimensionality_2021}. Thus, models have to learn a low-dimensional mapping of the operational space, explicitly incorporating information about the additional tasks such as for example plasma state disruptivity.

\subsection{Architecture}
The employed neural network model follows a general consistent architecture inspired by \cite{wei_dimensionality_2021}. Each model consisted of an encoder $q_\phi$ with parameters $\phi$ that maps the plasma state $\mathbf x \in \mathbb{R}^F$ to a latent variable $\mathbf z \in \mathbb{R}^L$. The encoded latent state $\mathbf z$ was further processed by $m$ parallel heads $q_{\psi_m}$ with parameters $\psi_m$ where each one solved a specific task. Let $\mathbf x^i_t$ be the plasma state at time $t$ of discharge $i$, then
    
\begin{equation*}
\quad \quad \quad \quad  \quad  \quad  \quad \mathbf{z}_t^i \sim q_\phi\left(\mathbf{z} \mid \mathbf{x}_t^i\right) \ ,
\end{equation*}

where the encoder $q_\phi$ is called state-based and parameterized the distribution over the possible latent variables as a symmetric Gaussian

\begin{equation*}
\quad \quad \quad  \quad  q_\phi\left(\mathbf{z} \mid \mathbf{x}_t^i\right)=\mathcal{N}\left(\mathbf{z} \mid \bm{\mu}_t^i,\left(\bm{\sigma}_t^i\right)^2 \mathbf{I}\right) \ .
\end{equation*}

The latent variable $\mathbf{z}_i^t$ was sampled from the posterior distribution using the reparameterization trick \cite{kingma_auto-encoding_2013}. The conditioning variable of the encoding network depends on the model type. If we let the encoder processes a sequence of plasma states

\begin{equation*}
\quad \quad \quad \quad \quad \quad \mathbf z^i_t\sim q_\phi(\mathbf z | \mathbf x^i_{t-k} , \ldots , \mathbf x^i_t) \ ,
\end{equation*}

the network is called sequence-based. Where state-based model used an MLP encoder, sequence-based models leveraged an autoregressive transformer encoder with a positional embedding relative to the start of the flat-top phase to learn temporal dependencies efficiently \cite{vaswani_attention_2017} as depicted by Figure \ref{fig:architecture}. To enhance model transparency, the latent space dimensionality was set to $L=2$ similar to prior work \cite{pau_first_2018, wei_dimensionality_2021}. The choice of the positional embedding on the start of the flat-top is not a constraint, but it allows, for this first demonstrative study, to relax the additional complexity due to the initial time-varying phase of the plasma current ramp-up.

\subsection{Tasks}
We focused on four tasks related to plasma state monitoring: disruptivity classification ($T_\mathrm{Disr}$), time-to-boundary regression  ($T_\mathrm{TTB}$), state reconstruction ($T_\mathrm{Rec}$) and smooth latent trajectory movements ($T_\mathrm{Mov}$). 

For disruptivity classification, we relied on a reference warning time similar to \cite{pau_machine_2019}, described in section \ref{s:expert-label}. It was used to generate binary labels for a disruptivity score $p_\mathrm{disr}$. We used binary cross-entropy $\mathcal{L}_\mathrm{BCE}$ as an objective function for this task.
To regress the time-to-boundary quantity, we use a boundary condition $T_B$ determined by considering plasma current ($I_p<1\mathrm{MA}$), ramp-down end ($t>T_\mathrm{RDE}$), time of the locked mode onset ($t>T_\mathrm{ML}$), and the time of disruption ($t>T_\mathrm{D}$). The relative time $t_\mathrm{B,rel}$ to $T_B$ served as the target variable. The quantity was clipped to a constant value of ten seconds if it was too large or if the shot experienced a stable ramp down. We utilize a symmetric mean absolute percentage error $\mathcal{L}_\mathrm{SMAPE}$ as an objective function for this task because it penalizes the scale of the error with respect to the scale of the target \cite{meade_long_1986}. Both $T_\mathrm{Disr}$ and $T_\mathrm{TTB}$ are supervised tasks, as they rely on supervised labels such as plasma disruptivity and time of disruption.

For state reconstruction, the input state $s_t$ is at the same time the target output. We minimize the mean squared error $\mathcal{L}_\mathrm{MSE}$ between the input plasma state and its reconstruction to maximize their similarity. We also introduced a smooth latent movement task that enforced smooth trajectory dynamics for consecutive latent variables of a discharge in the latent space. It was optimized by minimizing

\begin{equation*}
\hspace*{-.2cm} \mathcal{L}_\mathrm{Mov}=D_{\mathrm{KL}}\left[\mathcal{N}\left(\bm{\mu}^i_t, (\bm{\sigma}^i_t)^2\mathbf I\right) \| \mathcal{N}\left(\bm{\mu}^i_{t-1}, (\bm{\sigma}^i_{t-1})^2\mathbf I\right)\right]
\end{equation*}

which encouraged the model to project plasma states that are close in time, to coordinates that are close in the latent space. These two tasks are unsupervised because their objectives are based on plasma measurements and their temporal relationships without incorporating any external knowledge about the disruptivity of the plasma state.

 Additionally, we employed regularization in the form of a latent space prior similar to \cite{kingma_auto-encoding_2013}. The latent space alignment with a spherical prior

\begin{equation*}
\ \ \ \ \ \ \ \mathcal{L}_\mathrm{Prior}=D_{\mathrm{KL}}\left[\mathcal{N}(\mathbf 0,\mathbf I) \| \mathcal{N}\left(\bm{\mu}^i_t, (\bm{\sigma}^i_t)^2\mathbf I\right)\right]
\end{equation*}

encouraged the model to learn a continuous, spherical, zero-centered state distribution \cite{kingma_adam_2014}. Through this alignment, plasma state representations that lie on a continuous surface are favored, avoiding the formation of well separated clusters.  Individual loss terms were minimized jointly by optimizing over their weighted sum \cite{thung_brief_2018}.


\begin{table*}[h]
    \centering
    \setlength{\tabcolsep}{2pt}
    \resizebox{0.6\linewidth}{!}{
    \begin{tabular}{| l || c  | c | c | c |}
        \hline 
          & \multicolumn{2}{c |}{\textbf{Supervised Tasks}} & \multicolumn{2}{c |}{\textbf{Unsupervised Tasks}}\\
         \multirow{2}{*}{\textbf{Model}} & \boldmath $T_\mathrm{Disr}$ & \boldmath $T_\mathrm{TTB}$ & \boldmath $T_\mathrm{Rec}$ & \boldmath $T_\mathrm{Mov}$\\
         & $\uparrow$ AP [\%] & $\uparrow$ R$^2$ [\%]& $\uparrow$ R$^2$ [\%] & $\downarrow$ KLD [a.u.] \\
        \hline \hline
State-based & $73.5 \pm 3.2$   & \boldmath $45.4 \pm 4.8$  & \boldmath $72.0 \pm 1.9$  & $1.61 \pm 0.776$ \\
 \hline
Sequence-based  & \boldmath $73.8 \pm 6.9$  & $40.3 \pm 11.8$  & $68.3 \pm 1.3$  & \boldmath $0.03 \pm 0.003$ \\
        \hline
    \end{tabular}
    }
    \caption{Evaluation metrics for all learned tasks. Metrics were computed over all states $s_t$ for all test discharges and are reported as micro averages.}
    \label{tab:data-centric-results}
\end{table*}

\section{Dataset}
\label{s:dataset}

The dataset used in this work originates from ILW experimental campaigns performed at JET from 2011 to 2022. It contained measurements of 520 shots where 322 were disruptive and 198 non-disruptive. The discharges were selected to have a statistically significant representation of low and high-power/current plasma scenarios, including base key operation scenarios like the ITER baseline for instance \cite{giroud_towards_2014,garcia2021integrated,garzotti_IAEA_2023}. The resulting distributions of shot numbers and durations were non-uniform for both non-disruptive and disruptive shots. The dataset comprised various diagnostic, including "raw" physics quantities, typically used to describe the plasma states, as well as features engineered to encode specific physics mechanisms as described in \cite{pau_first_2018}. Metadata identifying shot phases, plasma regime transitions and precursor events (as reported in \cite{vries_survey_2011}) complement the characteristic times describing the main disruption phases, such as the time of disruption $T_\mathrm{D}$. All recorded signals are down- or resampled to the cycle time of the JET real-time network $f_s = 500\mathrm{Hz} = 2$ms \cite{pau_machine_2019}. 

\subsection{Reference Warning Time $T_\mathrm{predisr}$}
\label{s:expert-label}
In addition to the time of specifing events and the time of disruption $T_\mathrm{D}$, a special label titled $T_\mathrm{predisr}$ is defined to define the point in time at which the plasma state is considered to diverge from expected trajectory to become disruptive. It lies between the start of the pre-precursor and precursor phase, which are associated with the events leading to the disruption \cite{wesson_tokamaks_2011}. Derived from expert knowledge, the $T_\mathrm{predisr}$ label is based on the chain of events that ultimately result in a disruption, as described in various studies \cite{wesson_tokamaks_2011,vries_survey_2011}. It identifies the beginning of the unstable phase for all considered discharges, such that plasma states prior to it are considered stable and disruptive thereafter. The dataset is imbalanced, with a disruptive vs stable state ratio of $\eta=0.1725$ when using $T_\mathrm{predisr}$ to mark the start of the unstable phase. The automated labeling of the discharge, including the definition of discharges phases, plasma regime transitions and precursor events is handled thorugh the DEFUSE software supporting, among other things, the construction and validation of the EUROfusion Disruption Database \cite{defuse_2023}. DEFUSE allows the automated characterization of the chain of events leading to disruptions through physics-based and data-driven models. Statistical analysis and interactive validation tools allow to check the consistency of the automated detections. 

\begin{table}[b]
    \resizebox{\linewidth}{!}{
    \begin{tabular}{l l}
    \hline
    \textbf{Description} & \textbf{Acronym} \\
    \hline 
     Internal inductance & LI\ \\
    Safety factor at $r=0.95$ surface & Q95\ \\
    Normalized beta & BTNM\ \\
    Electron density peaking factor & HRTS\_Ne\ \\
    Electron temperature peaking factor & ECM1\_PF\\
    Core radiation peaking factor & BOLO\_HV\\
    Divertor radiation peaking factor & BOLO\_XDIV\ \\
    Normalized locked mode from saddle coils & ML\_norm\\
    Normalized $N=1$ mode amplitude & N1\_norm\\
    Normalized $N=2$ mode amplitude   & N2\_norm\\
    Greenwald fraction & GWfr\\
    Fraction of the effective input power  & P\_LHfr\\
    required to enter H mode &\\
    Fraction of total radiated power & Pfrac\_tot\ \\
    Error of plasma current w.r.t.\ reference  & Ip\_err\_norm\\
     plasma current & \\
     \hline
    \end{tabular}}
    \caption[Employed features]{List of the considered parameters.}
    \label{tab:features}
\end{table}

\begin{table*}[t]
    \centering
    \setlength{\tabcolsep}{2pt}
    \resizebox{\textwidth}{!}{
    \begin{tabular}{| l || c  | c | c | c  | c | c | c | c | c | c |}
        \hline 
         \multirow{2}{*}{\textbf{Method}} & \multicolumn{3}{c |}{\textbf{Network-only}} & \multicolumn{3}{c |}{\textbf{Network + ML}} & \multicolumn{3}{c |}{\textbf{Network + ML}} \\
         & $\downarrow$ MA [\%] & $\downarrow$ FA [\%]& $\uparrow$ SR [\%] & $\downarrow$ MA [\%] & $\downarrow$ FA [\%]& $\uparrow$ SR [\%] & $\downarrow$ MA [\%] & $\downarrow$ FA$^\prime$ [\%]& $\uparrow$ SR$^\prime$ [\%] \\
        \hline\hline
State-based               & \boldmath $2.39 \pm 1.5$   & $55.7 \pm 6.3$  & $72.75 \pm 2.9$  & $0.0 \pm 0.0$   & $55.7 \pm 6.3$  & $74.13 \pm 3.1$  & $0.0 \pm 0.0$  & $28.76 \pm 3.0$  & $85.37 \pm 1.7$\\ \hline
Sequence-based               & $9.54 \pm 2.4$   & \boldmath  $23.3 \pm 3.4$  & \boldmath  $84.02 \pm 1.2$  & $0.0 \pm 0.0$   & \boldmath  $23.3 \pm 3.4$  & \boldmath  $89.64 \pm 1.7$  & $0.0 \pm 0.0$   & \boldmath  $8.18 \pm 2.6$   & \boldmath  $96.2 \pm 1.2$\\
        \hline
    \end{tabular}}
    \caption{Disruption prediction performances of pure data-driven and hybrid detectors. The metrics FA$^\prime$ and SR$^\prime$ correspond to the adapted false alarm and success rates described in section \ref{ss:detection-rates}. Network-only reefers to using the network
disruptivity predictions only, triggering an alarm once a fixed threshold
is overstepped. Network + ML uses both network predictions and a mode lock indicator to trigger an alarm.}
    \label{tab:disruption-prediction-results}
\end{table*}

\subsection{Feature selection and preprocessing}
\label{s:fs-preprocessing}
 Table \ref{tab:features} reports the selected set of 14 features which was chosen to describe the plasma state ($F=14$), inspired by previous works on disruption prediction \cite{pau_machine_2019, pau_first_2018, wei_dimensionality_2021}. Most of the chosen features are inherently dimensionless, making them device agnostic. Additional discharge metadata was also extracted for preprocessing and target label generation.
 
 The domain of analysis described the window within which the models were supposed to monitor the plasma. The analysis domain was established from the flat-top start until the boundary condition $T_B$ was met, avoiding the non-stationary plasma current ramp-phase, where the plasma often has not yet reached the diverted configuration. A sample \& hold filter was employed for data imputation of NaN values within the defined analysis domain.
 After data consistency and integrity checks discarding the shots where part of the diagnostics were unavailable,  the remaining number of discharges was 430, including 266 disruptive and 164 non-disruptive plasma terminations. 
 
For quantitative evaluations the dataset was divided into multiple train (80\%), test (10\%), and validation (10\%) splits. Cross-validation over five splits was used to test for randomness in the training procedure and hyperparameter tuning. The splits were stratified over discharge duration. Features were normalized by subtracting the mean and dividing by their standard deviation, with statistics computed over the respective training sets.


\section{Results}

We trained state-based as well as sequence-based models for $10^5$ batch updates using the AdamW optimizer \cite{loshchilov_decoupled_2017}. Sequences were sampled with a length of 512 states, thus spanning approximately one second. The transformer model backbone was inspired by the GPT-2 architecture, but using a reduced model dimensionality  ($d_\mathrm{model}=192$), with only three attention heads ($h=3$) and twelve layers, ($K=12$) \cite{radford_language_2019} resulting in an order of magnitude less parameters. Disruptive states were resampled uniformly to account for label imbalance such that for each training batch $\eta \approx 1$. During training, the $T_\mathrm{predisr}$ label was augmented by a uniform random time shift ($\pm20$ms) to simulate the uncertainty of the label. Downsampling $f_s$ by a randomized factor allowed models to learn lower-frequency, long-term dependencies up to four seconds. 

Table \ref{tab:data-centric-results} reports on the evaluation performances on the considered learned tasks averaged over all state measurements of unseen test discharges. Scores of the tasks $T_\mathrm{Disr}, T_\mathrm{TTB}$ and $T_\mathrm{Rec}$ are on similar scales for both model types. They are also not very close to their optimal values. It is likely that by relaxing the strong constraints we imposed, such as $L=2$, we would be able to improve on those numbers. However, in this work, we are mainly interested in the transparency of the model's decision-making process. On the contrary for the $T_\mathrm{Mov}$ task, Table \ref{tab:data-centric-results} does report a significant difference between the state- and sequence-based approach, indicating that the latter does allow for smoother latent trajectory evolutions than the former. We also find rather large standard errors on the supervised tasks for both model types which is most likely due to data scarcity regarding samples within the disruptive phases.

\subsection{Disruption Prediction}

For a further comparison we assess the models' performances as disruption predictors. The networks are evaluated using two detection methods. The first one is referred to as \textit{Network-only} and uses the raw network prediction regarding plasma state disruptivity and triggers an alarm once a fixed threshold is overstepped. The second detection method is called \textit{Network + ML} and corresponds to a hybrid detector. It additionally leverages a locked mode onset event detection such that an alarm is triggered either if the disruptivity score prediction oversteps a threshold or if there was a locked mode onset event (similar to the basic control scheme described in \cite{pau_machine_2019}). In order to allow a comparison with other references in literature, we evaluate the performance of these detectors focusing on the missed alarm rate (MA), false alarm rate (FA) as well as success rate (SR).

\subsubsection{Detection Rates}
\label{ss:detection-rates}

\begin{table}[b]
    \centering
    \setlength{\tabcolsep}{2pt}
    \resizebox{\linewidth}{!}{
    \begin{tabular}{| l || c  | c | c |}
        \hline 
         \multirow{2}{*}{\textbf{Method}} & \multicolumn{3}{c |}{\textbf{Network + ML + Assertion Time}} \\
         & \ $\downarrow$ MA [\%] \ & $ \ \ \downarrow$ FA$^\prime$ [\%] \ \ & $\uparrow$ SR$^\prime$ [\%] \\
        \hline\hline
State-based                     & $0.0 \pm 0.0$   & $7.36 \pm 2.6$  & $96.68 \pm 1.2$\\
\hline
Sequence-based                & $0.0 \pm 0.0$   & \boldmath $6.13 \pm 3.0$   & \boldmath $97.6 \pm 1.1$\\
        \hline
    \end{tabular}}
    \caption{Disruption prediction performances with an assertion time window of 200ms.}
    \label{tab:disruption-prediction-results-assertion-time}
\end{table}

Using the Network-only approach, sequence-based models show a higher missed alarm rate but a lower number of false alarms as reported in Table \ref{tab:disruption-prediction-results}. When evaluating the hybrid detector, the performance improves by reducing the missed alarm rate to zero. However, their false alarm rate remains unchanged.

\begin{figure}[t]
    \centering
    \includegraphics[width=0.6\linewidth]{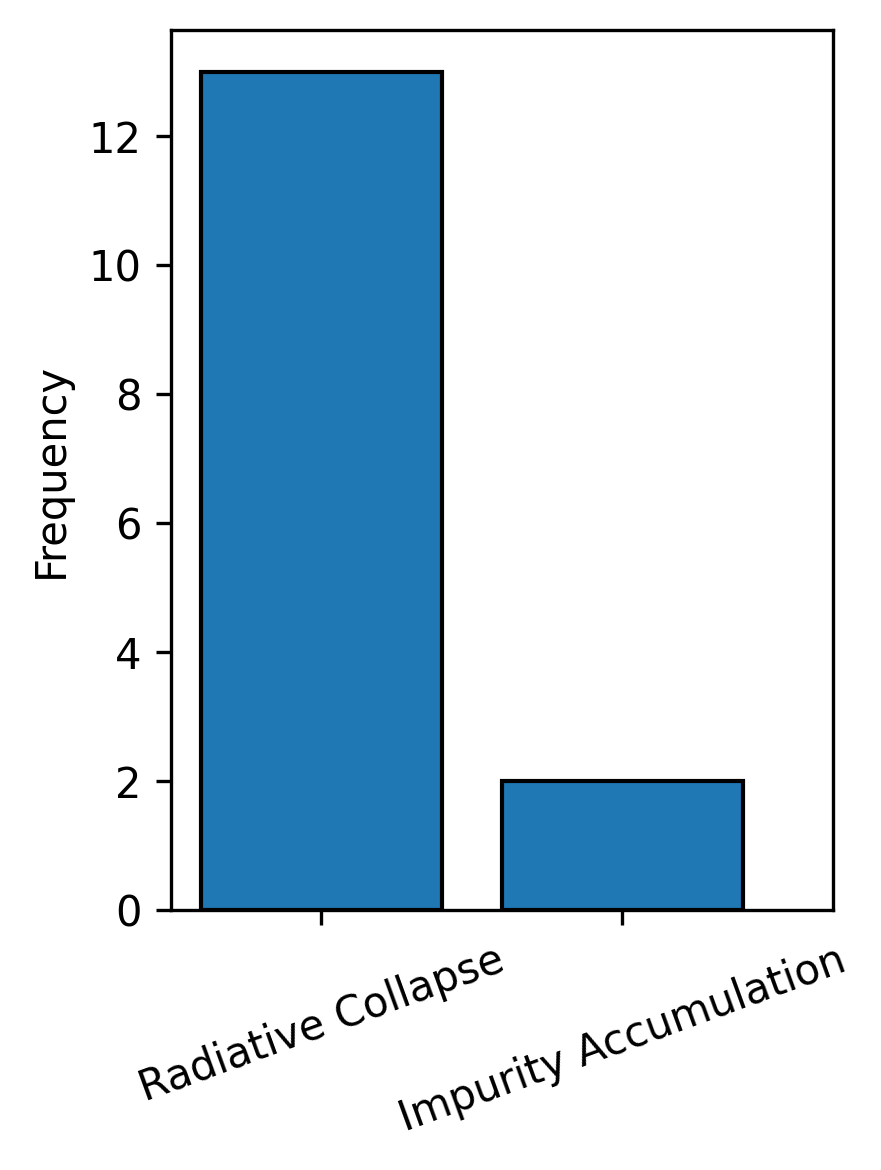}
    \caption{Instabilities found in a fixed window of $\pm200$ms around false alarms of sequence-based models.}
    \label{fig:false-alarms-instabilities}
\end{figure}

Upon further analysis, we found that the false alarms are correlated with labeled precursor events describing plasma instabilities. Figure \ref{fig:false-alarms-instabilities} shows that in a fixed time window of $\pm200$ms around false alarm predictions, there is a significant amount of labeled radiative collapse as well as impurity accumulation events. This raised questions about the definition of a generic false alarm for the disruption prediction task, in the presence of destabilizing events. The results presented in this paper call, among other things, for a more in-depth discussion about the requirements of the triggers in the context of disruption prevention. We refer the reader to the conclusions and the outlook for such a discussion. To account for this source of ambiguity, we reconsider the definition of the false alarm rate $\mathrm{FA}^\prime=\mathrm{FA} - \mathrm{FA}_\mathrm{instabilities}$ accounting for instabilities close by as correct detections. From $\mathrm{FA}^\prime$ we can also compute a success rate revisited accordingly $\mathrm{SR}^\prime$.

Considering these new metrics, Table \ref{tab:disruption-prediction-results} shows that the false alarm rate decreases and the success rate increases significantly. We can also see that for more than half of the false alarms of sequence-based models, there are destabilizing events close by. This indicates that such models might be well suited for disruption prevention, as they already learn to detect local instabilities without being explicitly trained for that purpose. Furthermore, the adapted success rate for sequence-based models is competitive with performances reported in prior work \cite{pau_machine_2019, wei_dimensionality_2021}.

\begin{figure}[t]
    \centering
     \begin{subfigure}[b]{0.4\textwidth}
         \centering
        \includegraphics[width=\linewidth]{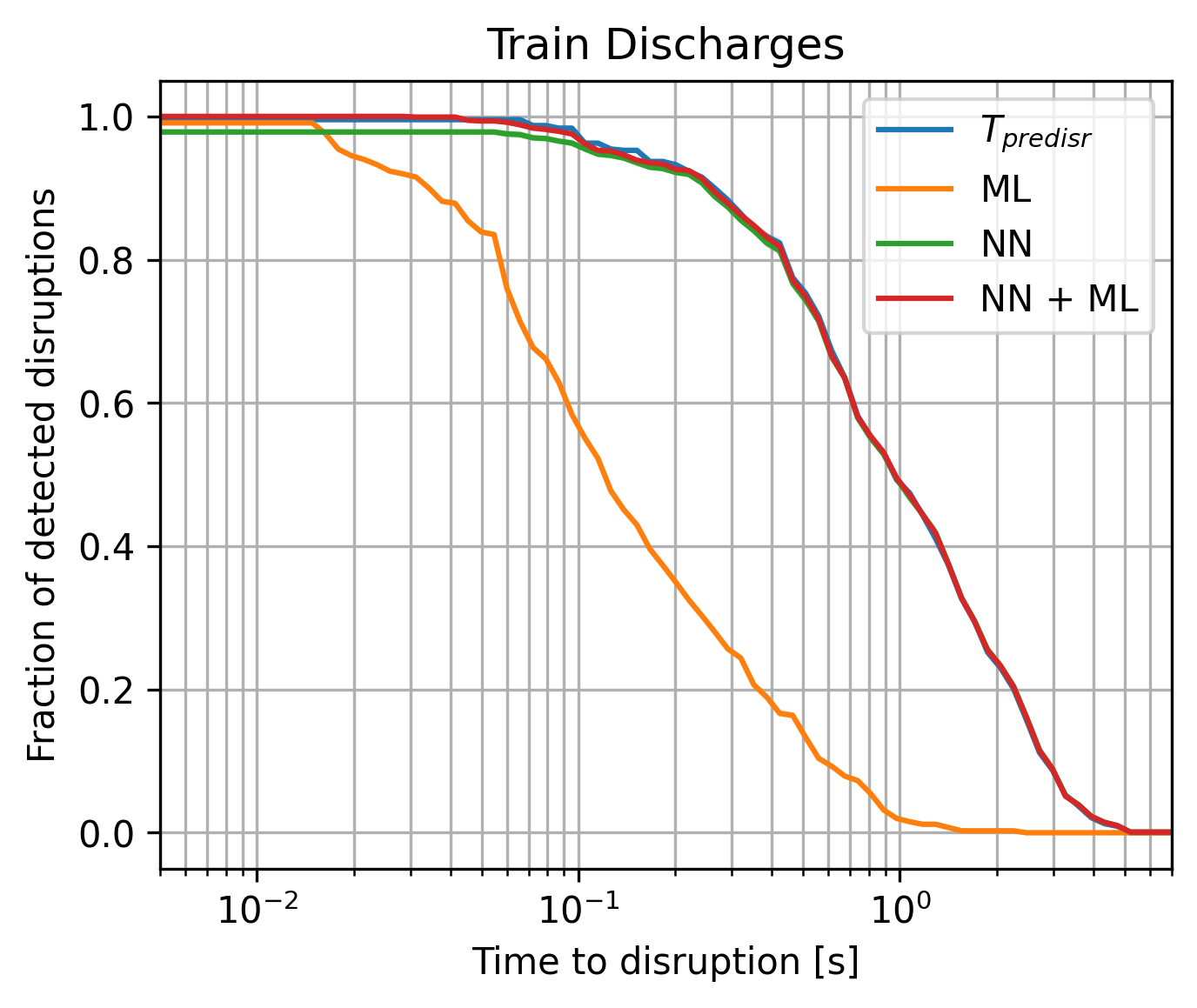}
        \caption{}
     \end{subfigure}
     \hfill
     \begin{subfigure}[b]{0.4\textwidth}
         \centering
        \includegraphics[width=\linewidth]{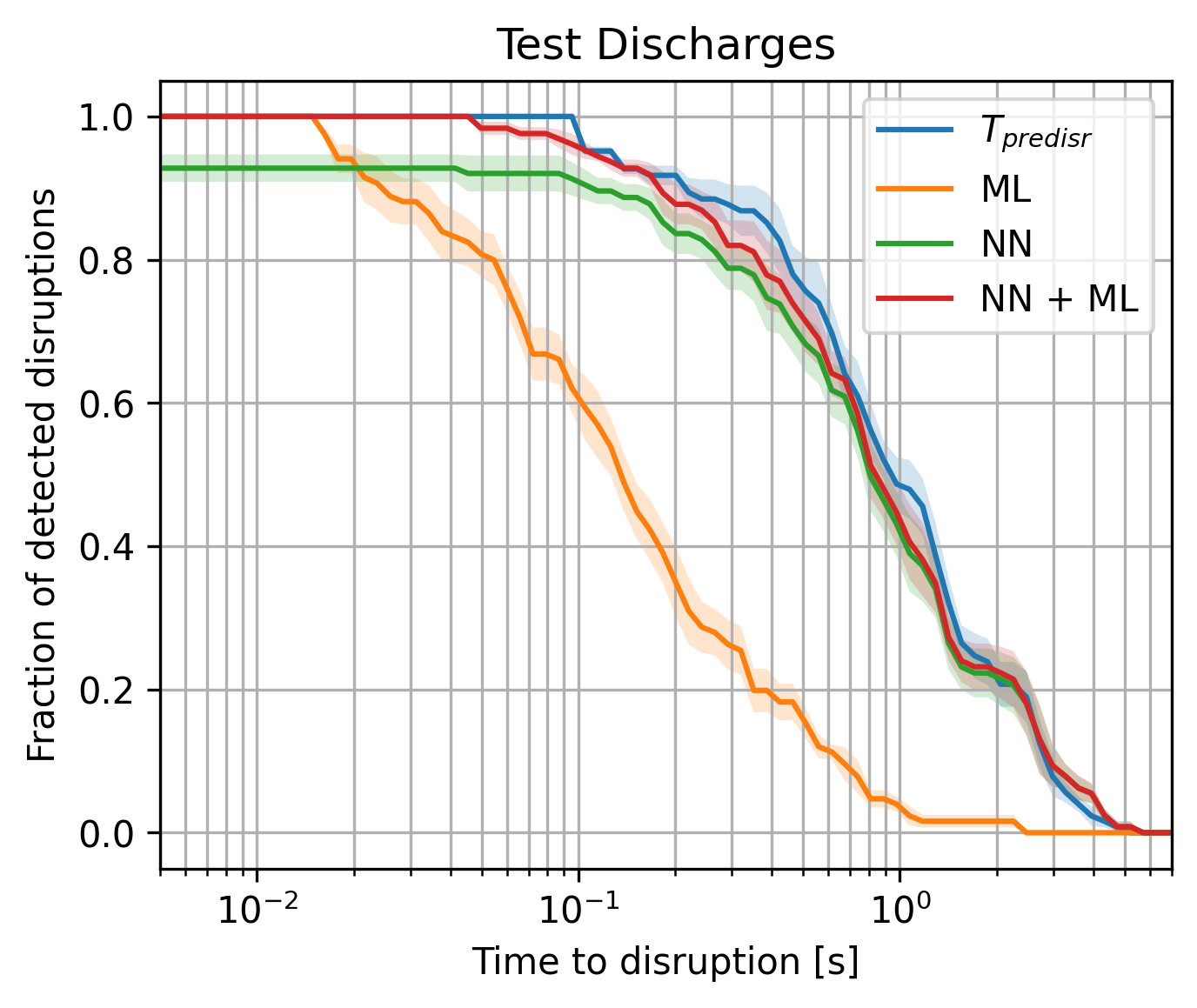}
        \caption{}
     \end{subfigure}
    \caption{Warning times for different detection methods using sequence-based models (a) on training discharges and (b) on unseen test discharges.}
    \label{fig:warning-times}
\end{figure}

\begin{table}[b]
    \centering
    \setlength{\tabcolsep}{2pt}
    \resizebox{0.9\linewidth}{!}{
    \begin{tabular}{| l || c | c |}
        \hline 
         \multirow{2}{*}{\textbf{Method}} & \textbf{Network-only} & \textbf{Network + ML} \\
         & $\uparrow$ MWT [s] & $\uparrow$ MWT [s]\\
        \hline\hline
State-based               &  $1.6 \pm 0.2$  &  $1.5 \pm 0.2$\\ 
\hline
Sequence-based              & $1.4 \pm 0.1$  & $1.3 \pm 0.1$\\
        \hline
    \end{tabular}}
    \caption{Mean warning times (MWT) of state and sequence-based models using a network-only or hybrid detector on unseen discharges.}
    \label{tab:mwt}
\end{table}

\subsubsection{Assertion Time}

To minimize the false alarm rate, one can additionally introduce an assertion time window similar to \cite{pau_machine_2019}, at the expense of potentially increasing the missed alarm rate. When using such a window, the prediction score must remain above a selected threshold for its entire duration, which introduces a hard-coded level of noise tolerance to the detector.

Table \ref{tab:disruption-prediction-results-assertion-time} shows that the introduction of an assertion window improved the success rate for sequence-based networks only slightly, suggesting that they already did learn to deal with measurement noise during training. On the other hand, for the state-based networks, we can observe a significant increase in terms of success rate, allowing them to achieve a similar performance as sequence-based models. Thus, the main difference between state-based and sequence-based models may be the ability of the latter to learn robustness to measurement noise during training.

\begin{figure*}[h]
    \centering
     \begin{subfigure}[b]{0.45\textwidth}
         \centering
        \includegraphics[width=\linewidth]{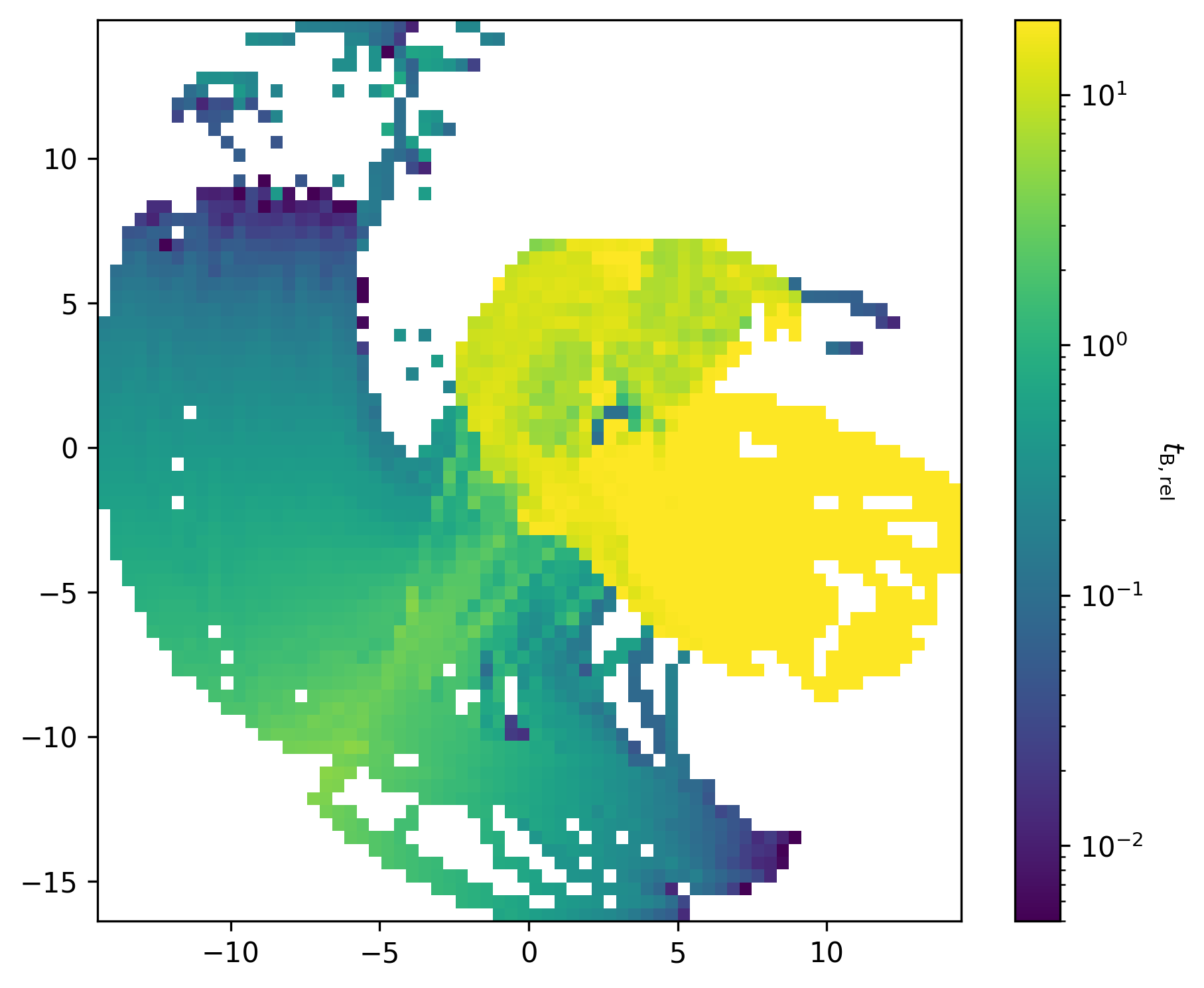}
        \caption{}
        \label{fig:operational-space-raw}
     \end{subfigure}
     \hfill
     \begin{subfigure}[b]{0.51\textwidth}
         \centering
        \includegraphics[width=\linewidth]{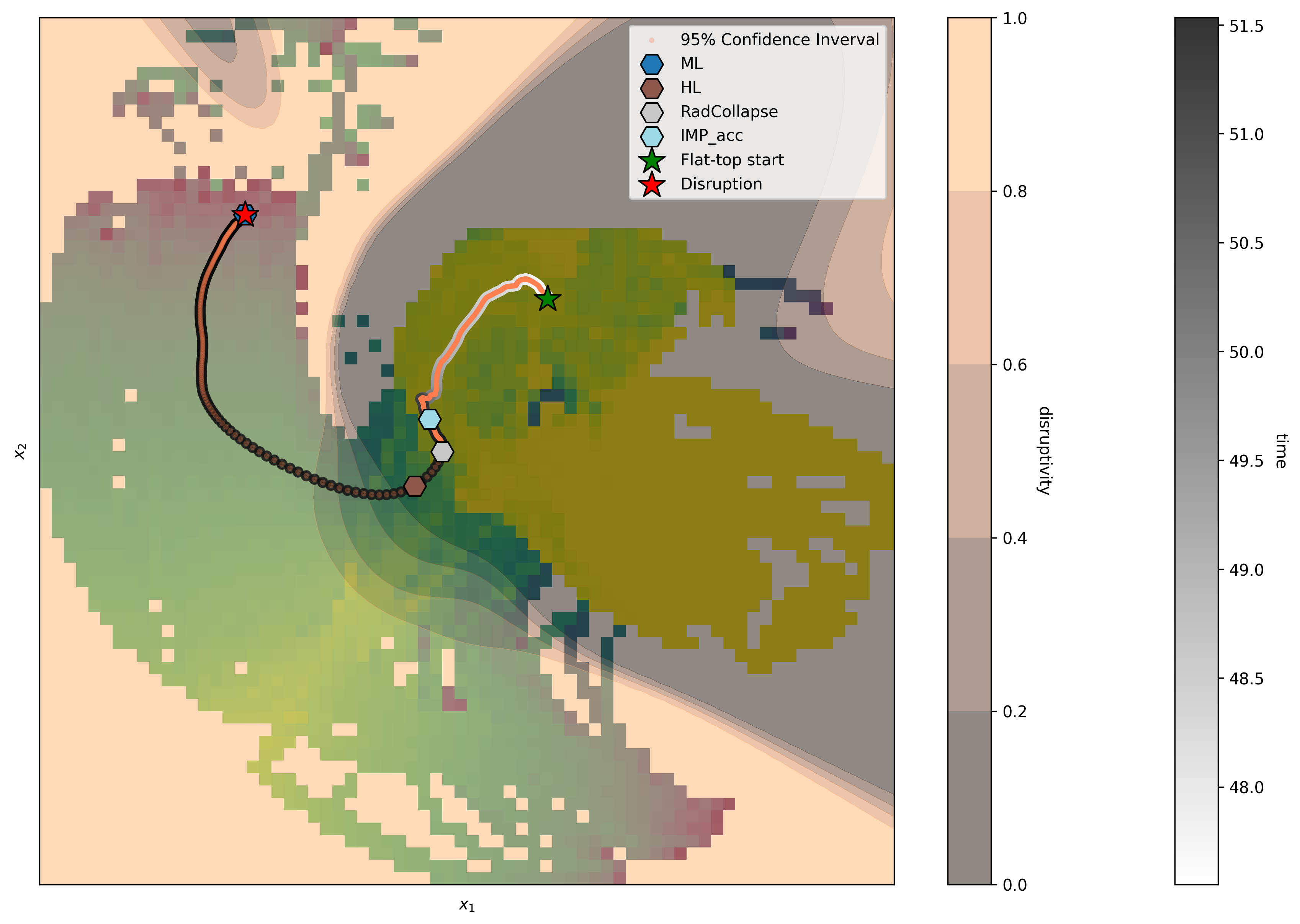}
        \caption{}
        \label{fig:tracking-single-shot}
     \end{subfigure}
    \caption{Visualization of the low dimensional projection of the training data. In (a) the color scale is log scaled and indicates the time to boundary quantity. For (b) the color scales indicate the predicted state disruptivity and the time variable of the shown discharge.}
    \label{fig:space-and-tracking}
\end{figure*}

\subsubsection{Warning Times}

Apart from the detection rates, the mean warning time (MWT) is another important performance indicator for a disruption predictor. This indicator is visualized by plotting the cumulative warning time distribution per detection method. 

Figure \ref{fig:warning-times} compares the cumulative warning time distributions of the ground truth $T_\mathrm{predisr}$ label, the locked mode onset event (ML),  the Network-only (NN), and the hybrid detection method (NN + ML) using sequence-based models. On the training discharges the network-only detector already seems to match the ground truth almost perfectly. However, for test discharges, we can observe that it does miss $\approx 10\%$ of the disruptions, whereas the hybrid detector does not. The hybrid detection method leads to almost perfect detections with a cumulative warning time distribution that closely resembles the ground truth on both the training as well as test discharges. This suggests that it is likely that not all causes leading to a locked mode event are described by the employed features at the given sampling frequency. Thus, there may be an additional set of features or a different, probably higher, sampling frequency that better describes the causes leading to a locked mode.

Table \ref{tab:mwt} reports the mean warning times for both considered approaches. State-based models show slightly higher MWTs, most likely due to their increased sensitivity to measurement noise, leading to some premature detections.

\section{Operational Space Mapping}

Figure \ref{fig:operational-space-raw} shows the latent space, learned by a sequence-based model on a grid map. The grid is colored by the log scaled mean, clipped time-to-boundary quantity $t_\mathrm{B,rel}$. The shape of the learned low-dimensional space is zero-centered and has spherical edges, which is likely due to the chosen spherical prior. We may already identify a vague disruptive boundary in the observed color scale. Furthermore, there appear to be two main and diverging paths along which the state can evolve towards the boundary condition. There also seems to be a small cluster of states in the top left corner that is not connected to the continuous state plane. 

\begin{figure*}[h]
    \centering
     \begin{subfigure}[b]{0.45\textwidth}
         \centering
        \includegraphics[width=.9\linewidth]{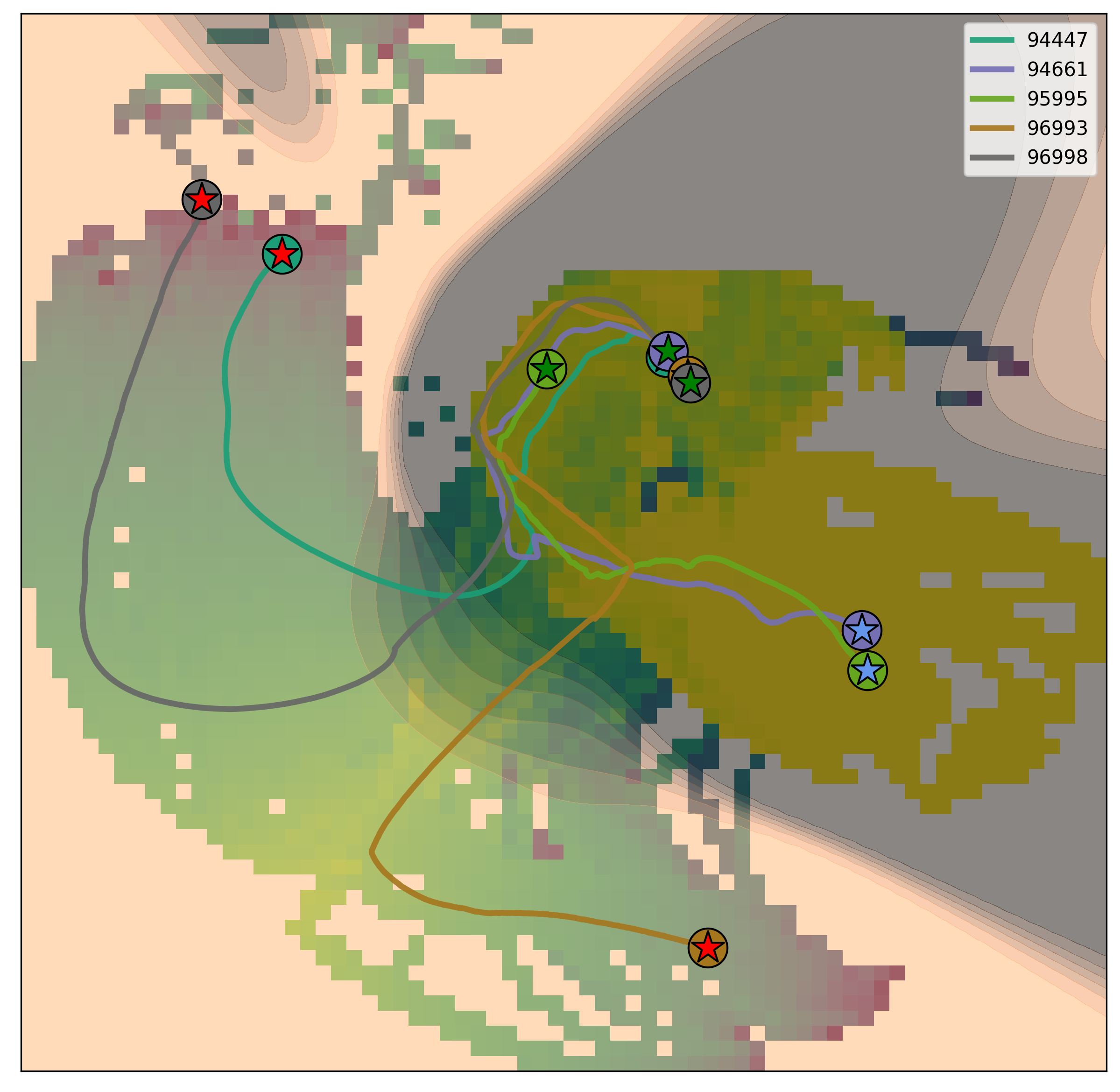}
        \caption{}
        \label{fig:tracking-multiple}
     \end{subfigure}
     \hfill
     \begin{subfigure}[b]{0.45\textwidth}
         \centering
        \includegraphics[width=.9\linewidth]{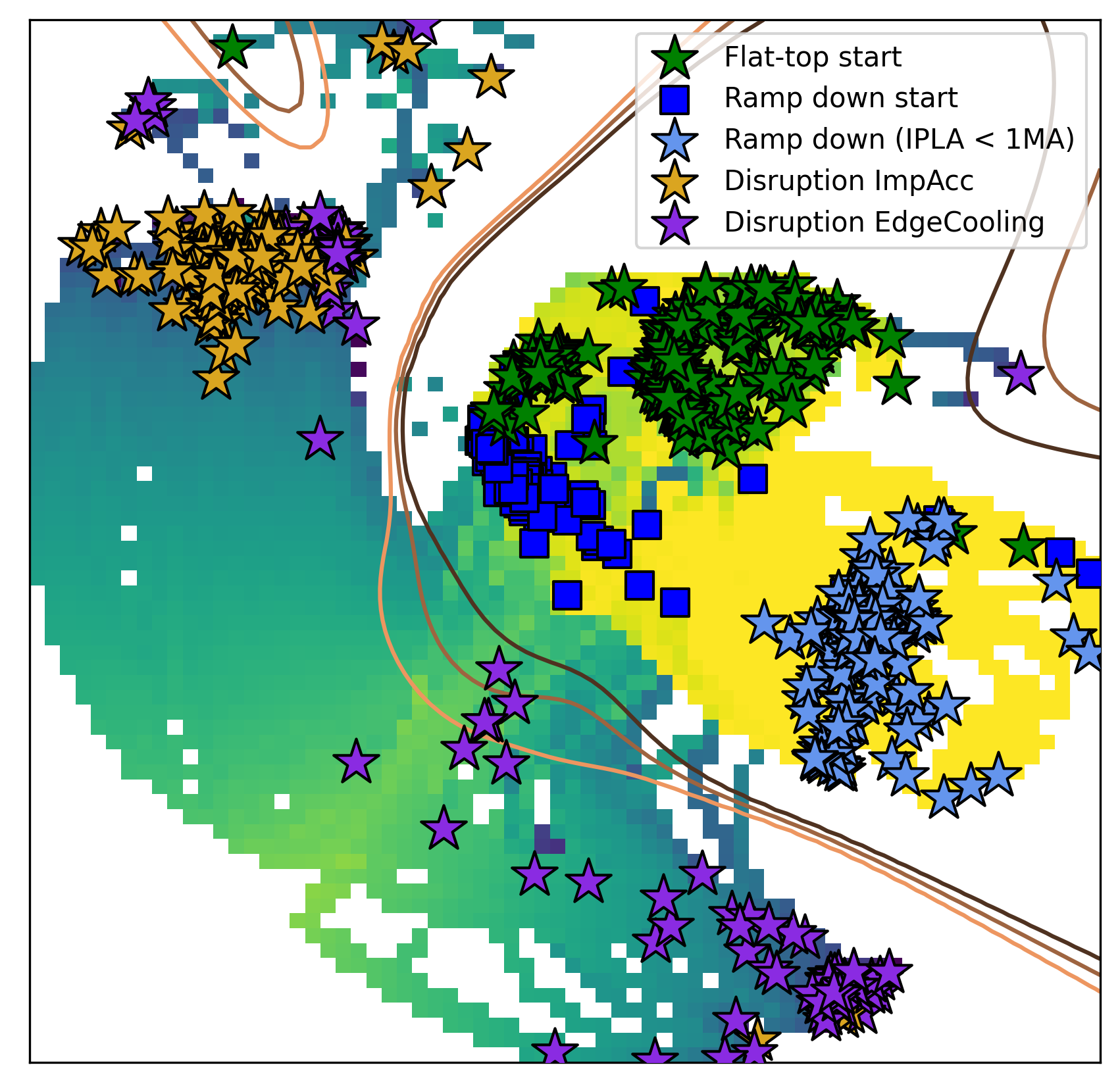}
        \caption{}
        \label{fig:heads-and-tails}
     \end{subfigure}
    \caption{(a) Multiple unseen discharges projected on the learned latent space where their start is called head and their end tail. (b) Visualization of all heads and tails of the training data.}
    \label{fig:heads-and-tails-tracking}
\end{figure*}

\subsection{Discharge Tracking}

The learned embedding space allows for the projection of measurements of unseen discharges onto it, enabling the visualization of the shot start (head), end (tail) and all states in between. Figure \ref{fig:tracking-single-shot} visualizes the trajectory of the unseen disruptive shot 94447 on the learned latent space. The trajectory starts at its head in the stable region and crosses at some point the learned disruptive boundary. Thereafter it follows a path along the contour lines of the $t_\mathrm{B,rel}$ quantity and encounters a locked mode at its tail. The trajectory also visualizes the 95\% confidence interval of a state measurement, which might indicate whether the plasma evolves slowly or undergoes a rapid change. The white-to-black color scale indicates the time variable, showing that the overall evolution of the trajectory is smooth as required by $T_\mathrm{Mov}$. Furthermore, the latent coordinates of the various labeled precursor events are also shown. The discharge seems to be crossing the learned disruptive boundary at the same time as it experiences a radiative collapse, corresponding to the definition of $T_\mathrm{predisr}$. 

A discharge tracking for multiple shots is depicted by figure \ref{fig:tracking-multiple}. The plot hints at a clustering of the heads and tails of the discharges, indicating that the latent space does incorporate some global dynamics. But trajectories do intersect on various occasions, implying that their evolution is still locally stochastic. However, all trajectories show smooth dynamics which may indicate that their local dynamics are predictable given their past evolution.

\begin{figure*}[h]
    \centering
     \begin{subfigure}[b]{0.45\textwidth}
         \centering
        \includegraphics[width=\linewidth]{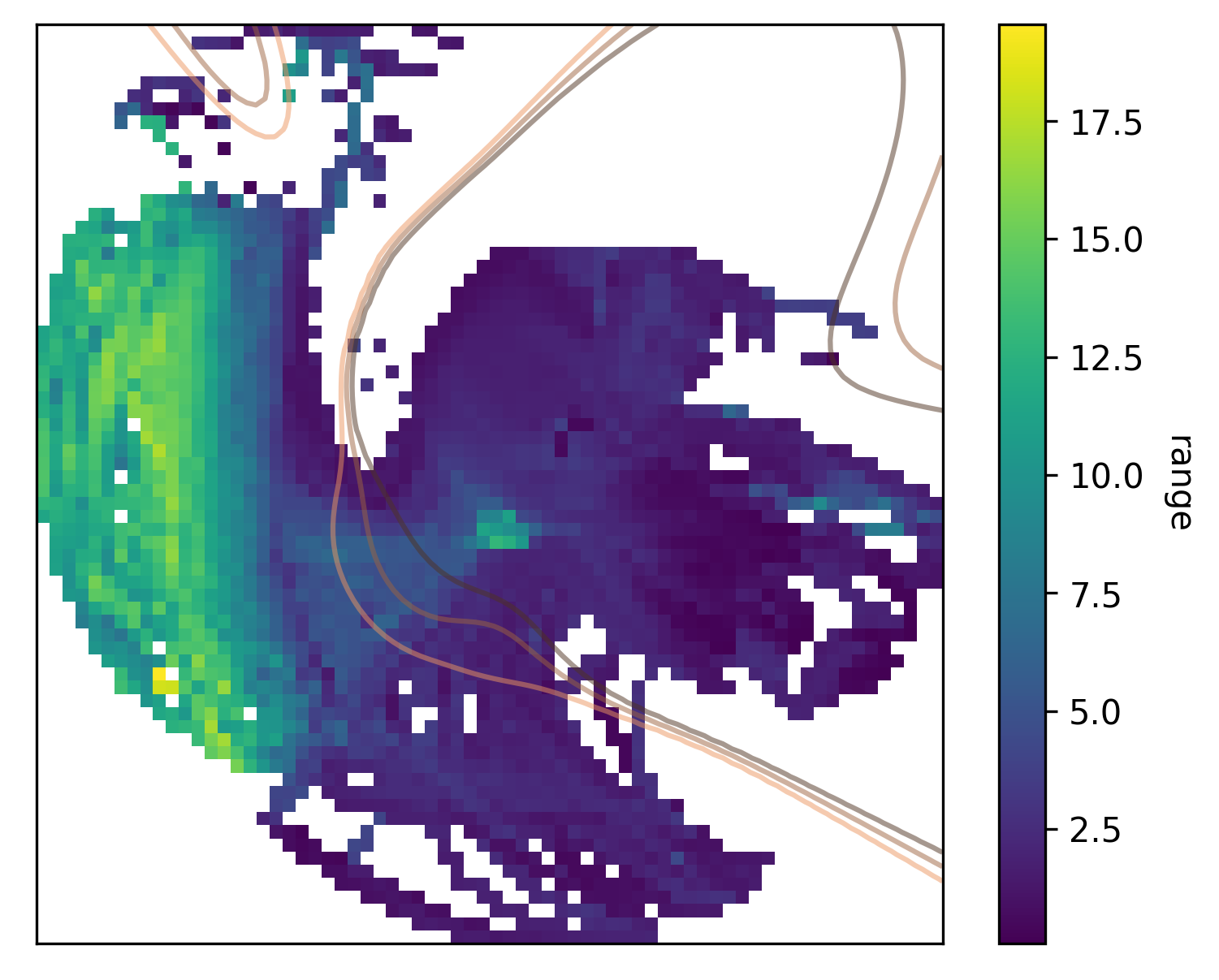}
        \caption{}
        \label{fig:feature-bolo-hv}
     \end{subfigure}
     \hfill
     \begin{subfigure}[b]{0.45\textwidth}
         \centering
        \includegraphics[width=\linewidth]{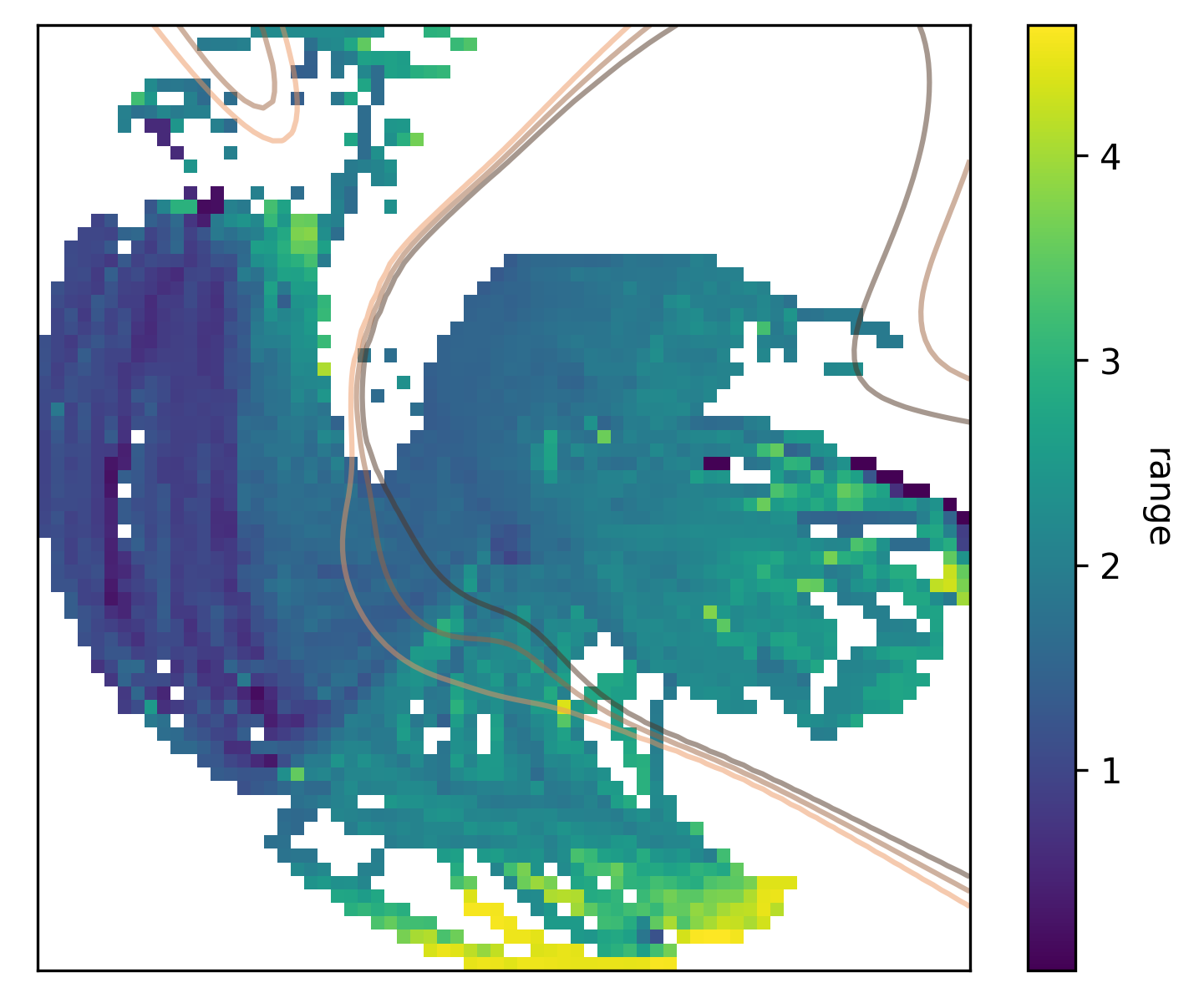}
        \caption{}
        \label{fig:feature-ecm1-pf}
     \end{subfigure}
    \caption{Component plane representation of (a) the peaking factor of the radiation and (b) the peaking factor of the temperature. The color bars show the scale of the dimensionless variables.}
    \label{fig:feature-distributions}
\end{figure*}

\subsection{Heads And Tails}
By computing a discharge tracking for all discharges and visualizing only the start and end state for each shot, one gets a heads and tails plot as shown in Figure \ref{fig:heads-and-tails}. The plot also shows the ramp-down start states of stable discharges. There seem to be two flat-top start clusters that suggest two distinct ramp-up strategies, while only a single ramp-down end cluster is revealed. These clusters correspond to three local optima of minimal disruptivity within the stable region. 
Disruptive tails are further divided into impurity accumulation and edge cooling macro classes, which correspond to different physical disruptive paths. The two diverging branches in the disruptive region strongly correlate with those two macro classes.

Visualizing the stable ramp-down start states allows the segmentation of the learned latent space into five regions, including a stable flat-top, stable ramp-down, transition, fast disruptive, and slow disruptive region. The stable flat-top region is characterized by the area enclosed by green stars and blue squares. A plasma evolving within this area may be unlikely to be unstable. The stable ramp-down operational space is enclosed by the blue squares and blue stars, and discharges in this region were likely to have had a stable flat-top and were ramping down their plasma current in a controlled manner. The transitional region, located between the prior two regions and the disruptive boundary, might correspond to an area where instabilities are building. The fast disruptive region is the shorter branch converging towards the lower right corner, indicating faster dynamics. The slow disruptive region corresponds to the larger branch converging towards the upper left corner, allowing for longer and more diverse dynamics.

A more detailed analysis revealed some flat-top start states outside their main clusters and even close to the disruptive boundary. Those cases were found to correspond to discharges with a rather short duration and were thus projected to those strange coordinates due to the movement penalty as an overfitting artifact. 

The disruptive tails also witness some curious patterns, with a scattering of tails in the fast disruptive region close to the disruptive boundary. All scattered tails were found to have a mode lock precursor event, while disruptions without such a precursor are clustered within the slow disruptive region. Furthermore, the scattered tails are all edge cooling disruptions with fast dynamics ($T_\mathrm{D}-T_\mathrm{predisr}<200$ms). Thus increasing the sampling frequency of the analyzed sequences could potentially help the network to project those tails into the correct disruption cluster at the tip of the diverging branch. Lastly, the scattered cluster in the upper left corner of the slow disruptive region was found to exclusively contain disruptive discharges that experienced a large plasma current error at the boundary condition.

\subsection{Marginal Feature Distributions}

Figure \ref{fig:feature-distributions} reports the component planes of two interesting input parameters used for the mapping. They reflect the marginal feature distributions and allow for uncovering patterns in the data. The radiation peaking factor BOLO\_HV shows high values in the slow disruptive region. Said region mostly corresponds to impurity accumulation disruptions in Figure \ref{fig:heads-and-tails} for which one would expect high values for the radiation peaking factor as also found by \cite{pau_machine_2019}. Figure \ref{fig:feature-ecm1-pf} shows that the temperature peaking factor spikes at the tip of the diverging branch in the fast disruptive region. That location corresponds to an edge cooling disruption cluster as seen in Figure \ref{fig:heads-and-tails} which also is expected and was also observed in \cite{pau_machine_2019}. These feature distributions can be interpreted as a learned global feature importance regarding state disruptivity and disruption macro class type.

\begin{figure}[t]
    \centering
    \includegraphics[width=0.8\linewidth]{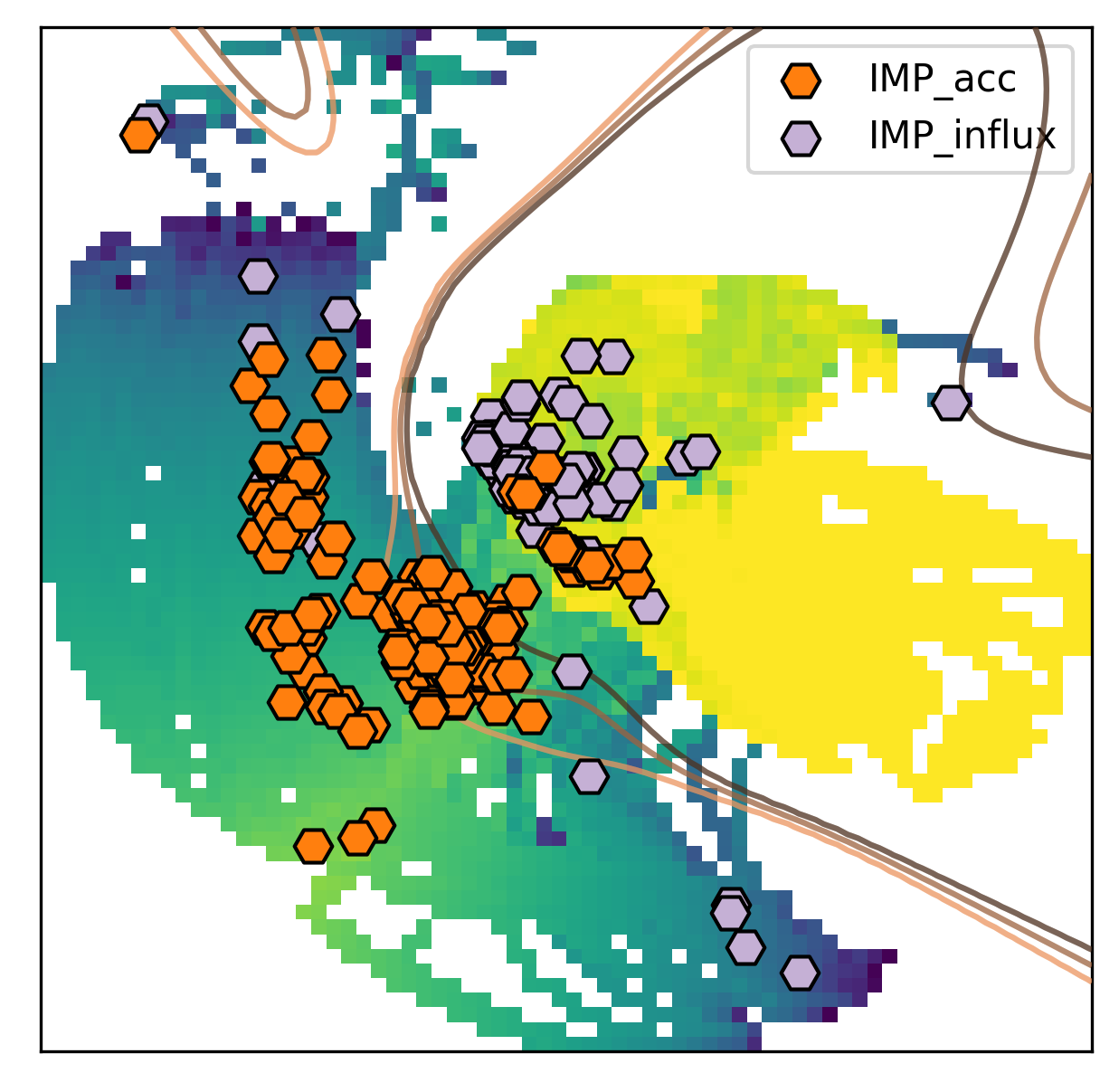}
    \caption{Distribution of events preceding  impurity accumulation disruptions. }
    \label{fig:events-distribution}
\end{figure}

\subsection{Precursing Events}
For each discharge, the precursor events as described by \cite{vries_survey_2011} were labeled semi-automatically by an expert. Figure \ref{fig:events-distribution} displays the distribution of two prominent events, by projecting the closest measurement in time into the latent space.

Both events are precursors to an impurity accumulation disruption. They do exhibit a temporal coherence in the mapping, where the impurity influx seems to be happening in the stable region and the accumulation in the transitional and disruptive area. This suggests that the model has learned about the temporal causality of these events. Furthermore, it also hints at the fact that they are not likely to be final precursors. The separation of the two events is not expected to be perfect, as an impurity influx not always results in an impurity accumulation.


\section{Conclusions}

The presented study has demonstrated the capability of neural networks to describe in a meaningful way complex high-dimensional patterns into a lower-dimensional representation, unveiling operation boundaries separating different plasma regimes. The resulting 2D latent space representation (similarly to what was done in  \cite{pau_machine_2019}) has successfully clustered the two main patterns characterizing JET disruptions, namely the core radiative collapse due to the accumulation of high-Z impurities, and edge radiative collapses characterized by the contraction of the plasma current channel, both leading to MHD unstable scenarios. A fundamental improvement with respect to the previous work is the capability of this model of learning the temporal dependencies hidden in time series data. This capability allows the learned latent space to encode plasma dynamics, describing more consistently plasma trajectories across different regimes, as well as characteristic patterns in the proximity to disruption boundaries. This aspect is of paramount importance in tasks such as plasma state monitoring, where the temporal dependencies play a key role in detecting and predicting changes in relevant operational parameters, as well as to further understand the mechanisms that lead to those changes.

When evaluated on the task of disruption prediction, the model achieved competitive success rates when considering nearby instabilities and including a mode lock event detection. The resulting warning time distribution closely resembled the ground truth, showing its potential for the implementation of disruption avoidance strategies. Furthermore, discharge tracking established smooth trajectory evolutions within the latent space, hinting at the fact that their dynamics might be predictable. 

Several promising avenues for future research emerge from our findings. The latent trajectory movements show curious patterns for disruptive discharges. Using methodologies such as integrated gradients \cite{sundararajan_axiomatic_2017}, one could attribute the predicted coordinate change to the input features, which could potentially uncover new information about the disruptive phenomenon. Furthermore, the adoption of self-supervised approaches from other domains could allow leveraging larger, unlabeled data sources for pretraining, potentially discovering more general and meaningful latent representations \cite{baevski_wav2vec_2020}. Finally, the observed smooth trajectories in the latent space may indicate a potential for learning latent plasma dynamics. Generative, autoregressive neural networks showed promising results for that task in other domains such as audio or natural language processing \cite{dhariwal_jukebox_2020, brown_language_2020}. A learned model of latent plasma dynamics could facilitate computer-aided scenario optimization. It would allow searching the space of possible plasma scenarios using automatized algorithms, significantly reducing the experimental time required to reach the necessary plasma conditions in a reactor. 

In conclusion, this paper showed that the applied method was able to learn a low-dimensional representation of plasma states, featuring consistent semantics regarding disruptivity, global dynamics, and locally smooth trajectory evolutions. By providing insights into the dynamics of different experiments,  plasma state monitoring allows researchers to optimize experimental scenarios and thus has the potential to further accelerate fusion research.

\section*{Acknowledgments}
This work has been carried out within the framework of the EUROfusion Consortium, partially funded by the European Union via the Euratom Research and Training Programme (Grant Agreement No 101052200 — EUROfusion). The Swiss contribution to this work has been funded by the Swiss State Secretariat for Education, Research and Innovation (SERI). Views and opinions expressed are however those of the author(s) only and do not necessarily reflect those of the European Union, the European Commission or SERI. Neither the European Union nor the European Commission nor SERI can be held responsible for them.

\section*{References}
\bibliography{iopart-num}

\end{document}